\documentclass[fleqn,twoside]{article}
\usepackage{espcrc2}
\usepackage{graphicx}
\usepackage[figuresright]{rotating}

\title{Performance of the CDF Miniplug Calorimeters}

\author{Michele Gallinaro \address{The Rockefeller University,\\ 
	1230 York Avenue, Box 188, \\ New York, NY 10021, USA}%
 }
       
\begin{document}

\begin{abstract}

Two Miniplug calorimeters, designed to measure the energy and lateral position of particles 
in the forward pseudorapidity region of 3.6~$<|\eta |<$~5.1, have been installed 
as part of the CDF upgraded detector for Run~II at the Tevatron.
Proton-antiproton beams are colliding at $\sqrt{s}$=1.96~TeV. 
One year after installation, Miniplug detector performance and first results are presented.

\vspace{1pc}
\end{abstract}

\maketitle

\section{INTRODUCTION}

During Run~I, which started in 1992 and lasted until the end of 1995, the CDF experiment 
collected a large data sample which has been extensively studied. 
Considerable knowledge on diffractive physics phenomena was gained
using those data (see, for example, Ref.~\cite{hcp}).

The Run~II physics program at the Tevatron Collider started in November 2001.
Protons and antiprotons are colliding at an energy of $\sqrt{s} = 1.96$~TeV,
with a typical instantaneous luminosity of ${\cal L}\approx 2 \div 3\cdot 10^{31}$cm$^{-2}$sec$^{-1}$.
Both the CDF and the D\O~experiments underwent major upgrade projects to
improve their detector capabilities.
Among these, the Forward Detectors upgrade project at CDF will enhance 
the sensitivity for hard diffraction and very forward physics during Run~II.

The signature of a typical diffractive event in $p\bar{p}$ collisions is 
a leading proton or anti-proton 
and/or a region at large pseudorapidity with no particles, also known 
as {\it gap} region.
In order to detect such events, forward regions in pseudorapidity are extremely important.
The diffractive physics topics to be addressed in Run II include 
studies of soft and hard diffraction,
searches for centauros and disoriented chiral condensates, and forward jet
production. 

The Forward Detectors include the {\it Roman Pot Spectrometer} 
fiber tracker detectors to detect leading antiprotons,
a set of {\it Beam Shower Counters} (BSCs)
installed around the beam-pipe at three (four) locations along the $p$($\overline p$)
direction to tag rapidity gaps at 5.5~$<|\eta |<$~7.5, 
and two forward {\it MiniPlug} (MP) calorimeters covering the pseudorapidity
region 3.6~$<|\eta |<$~5.1.
All the above detectors have been installed, are now fully integrated with 
the rest of the CDF detector, and are presently collecting data.

In the following sections, the MP detectors~\cite{mp_nim} and their performance will be presented. 
The other Forward Detectors are discussed elsewhere~\cite{fd_proposal,rio} 
together with the main goals of the physics program.

\section{DETECTORS}

The program of hard diffraction and very forward physics for Run~II benefits from
two forward MP calorimeters 
designed to measure the energy and lateral position of both
electromagnetic and hadronic showers. 
The MPs can detect both charged and neutral particles.
\noindent
They extend the pseudorapidity
region covered by the Plug calorimeters, which is 1.1~$<|\eta|<$~3.5. 
\begin{figure}[htb]
\includegraphics*[width=\hsize]{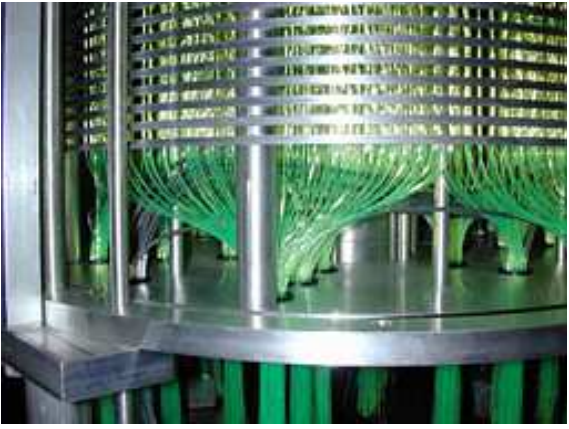}
\caption{\label{fig:mp_fibers} Fiber routing in the Miniplug.}
\end{figure}
The MP and Plug calorimeters can measure the width of the rapidity gap(s)
produced in diffractive processes and will allow extending Run~I studies of
the diffractive structure function to much lower values of the fractions $\xi$,
where $\xi$ is the momentum of the proton carried by the pomeron. 
The low $\xi$ values can be measured from the size of the rapidity gap
region using information from both BSCs and MPs.

\begin{figure}[htb]
\includegraphics*[width=\hsize]{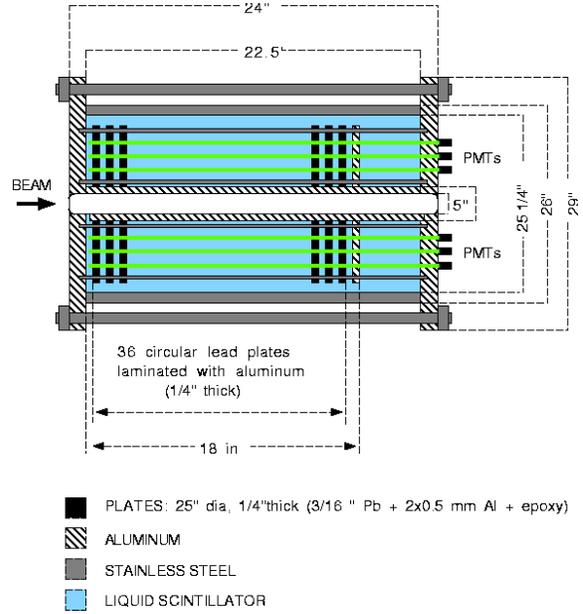}
\vspace*{-1.0cm}
\caption{\label{fig:mp_side}
Side view of a Miniplug (not to scale).}
\end{figure}

The MPs consist of alternating layers of lead plates and liquid scintillator 
read out by {\it Wave-Length Shifting}~(WLS) fibers (Fig.~\ref{fig:mp_fibers}).
The WLS fibers are perpendicular to the lead plates and parallel to the 
proton/anti-proton beams, in a geometry where towers are formed by combining
the desired numbers of fibers
and are read out by {\it Multi-Channel PhotoMultipliers}~(MCPMTs).
The 16-channel R5900 MCPMTs have been produced by Hamamatsu with a quartz window 
which significantly improves the radiation hardness.
The MP has a ``towerless'' geometry and has no dead regions due to the 
lack of internal mechanical boundaries.
Each MP is housed in a cylindrical steel barrel $\boldmath 26''$ in diameter and has a 
$\boldmath 5''$--hole concentric with the cylinder axis to accommodate the 
beam-pipe (Fig.~\ref{fig:mp_side}).
The active depth of each MP is 32 radiation lengths and 1.3 interaction lengths.
The ``short'' hadronic depth does not allow a large lateral spread of the 
showers, thereby facilitating the determination of the shower position and particle counting.

\begin{figure}[htb]
\includegraphics*[width=\hsize]{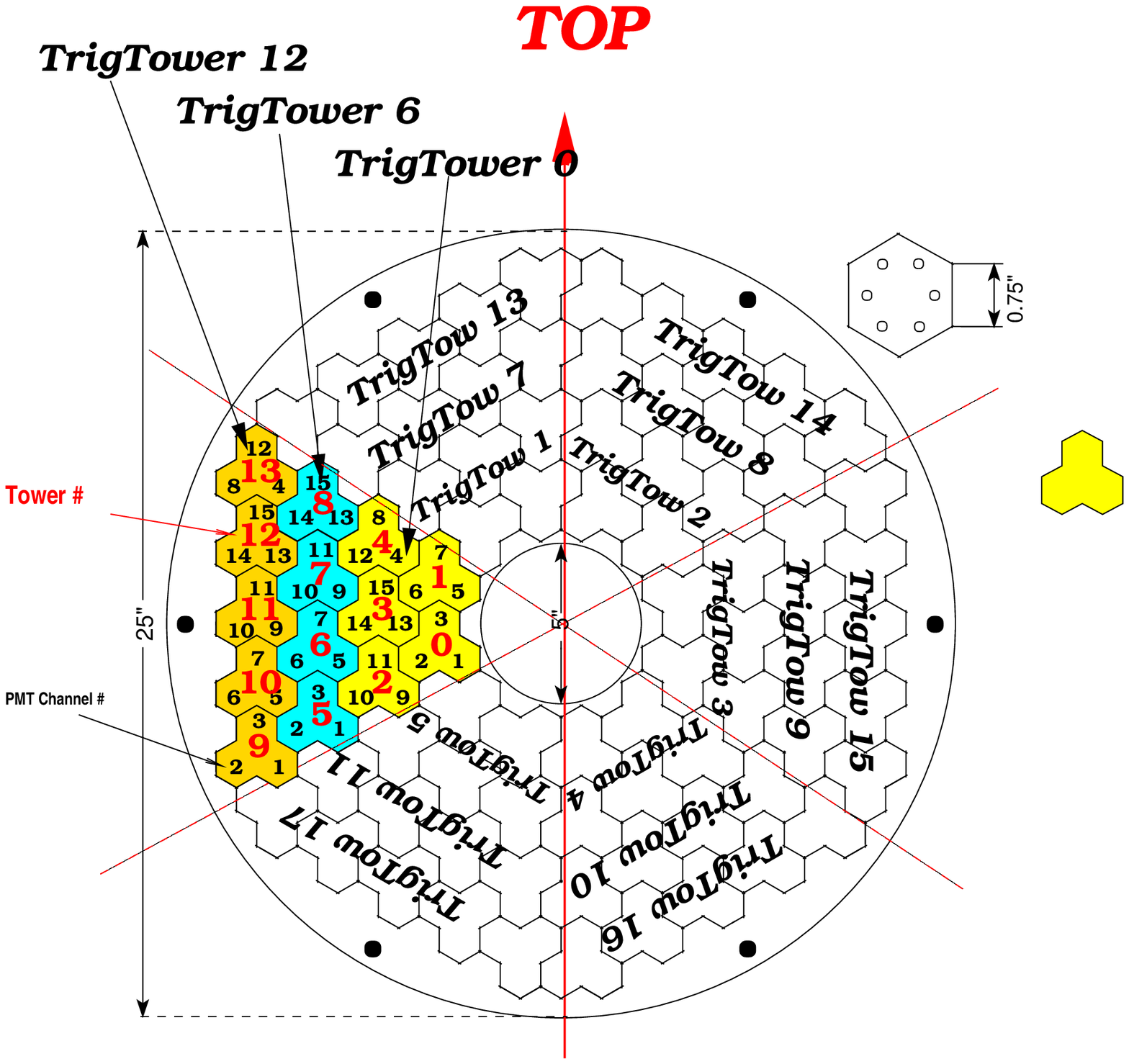}
\vspace*{-1.5cm}
\caption{\label{fig:mp_tower_geometry} 
Tower geometry of the East Miniplug calorimeter (viewed from the interaction point).}
\end{figure}

The design is based on a hexagon geometry. Uniformly distributed over each
plate, holes are conceptually grouped in hexagons and each hexagon has six holes.
A WLS fiber is inserted in each hole. The six fibers of one hexagon are grouped
together and are viewed by one MCPMT channel.
The MCPMT outputs are added in groups of three to form 84 calorimeter towers in order to
reduce the costs of the readout electronics.
The tower geometry is organized in four concentric circles around the beam-pipe
(Fig.~\ref{fig:mp_tower_geometry}).

The entire MCPMT can also be read out through the last dynode output,
indicated as {\it TrigTower} in Figure~\ref{fig:mp_tower_geometry},
to provide triggering information.
Each MP has a total of 18 trigger towers, arranged in three rings,
the {\it inner}, the {\it middle} and the {\it outer} ring. 
This allows triggering on different pseudorapidity ($\eta$) regions, either for events with a 
{\it gap} region or for events with large energy clusters.
An additional clear fiber carries the light from a calibration LED to each MCPMT pixel.
The LED allows a first relative gain calibration to equalize the MCPMT gains.
It also allows periodical monitoring of the MCPMT response.

\begin{figure}[htb]
\includegraphics*[width=\hsize]{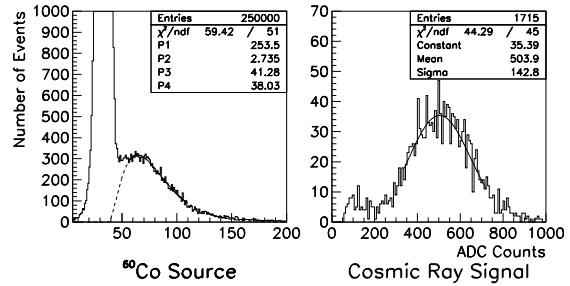}
\vspace*{-1.5cm}
\caption{\label{fig:cosmic_test}
Cosmic ray test of {\it Tower} \#7 of the east Miniplug.
$^{60}$Co source signals; {\it P3} parameter corresponds to the
pulse height of a single photoelectron (left).
Cosmic ray spectrum after an isolation cut fitted to a
Gaussian distribution (right).}
\end{figure}

Cosmic ray muons were used to test one 60$^\circ$-wedge of the East MP.
In this test, the cosmic ray trigger fired on a 2-fold coincidence 
of scintillation 
counter paddles located on top and at the bottom of the MP vessel,
placed with the towers pointing upward.
The outputs from Towers~\#5,~6,~7 and~8 and from Trigger Towers~\#0,~6 and~12 
(Fig.~\ref{fig:mp_tower_geometry}) were read out.
An energy isolation cut selected only those muons which went
through the entire length of the central Trigger Tower (\#6) and vetoed 
on the signals from the neighboring Trigger Towers (\#0 and 12).
The single photoelectron response for Tower \#7 was measured using 
a randomly gated signal from a $^{60}$Co source 
(Fig.~\ref{fig:cosmic_test}).
The single tower response to a {\em Minimum Ionizing Particle} was found to be
approximately 120 photoelectrons, exceeding the design specification.

\begin{figure}[htb]
\includegraphics*[width=\hsize]{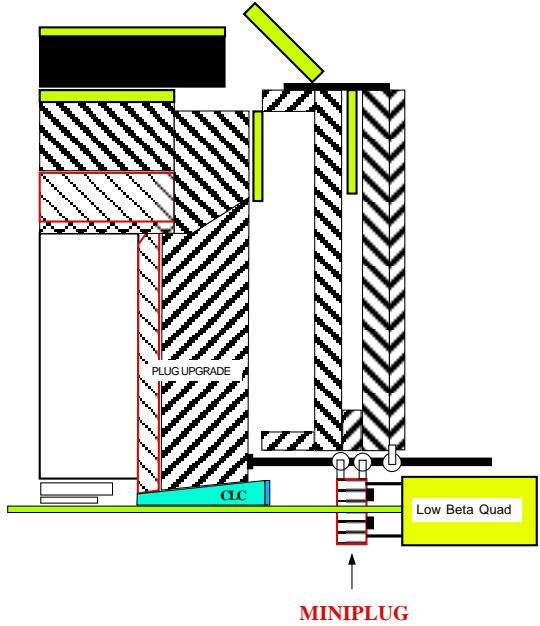}
\vspace*{-1.4cm}
\caption{\label{fig:mp_at_cdf}
Schematic drawing of a quarter view of the CDF detector showing a
Miniplug calorimeter installed inside the toroids (not to scale).}
\end{figure}

The MPs have been installed along the beam-pipe within the hole of
the muon toroids at a distance of 5.8~m from the center of the CDF detector
(Fig.~\ref{fig:mp_at_cdf}).
Since June 2002, the MP detectors are fully instrumented and collecting data.

\section{CALIBRATION AND RESULTS}

\begin{figure}[htb]
\includegraphics*[width=\hsize]{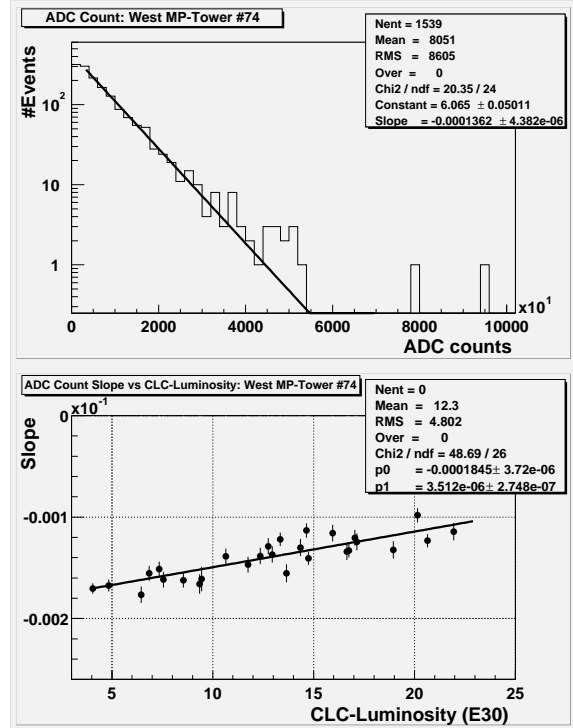}
\vspace*{-1cm}
\caption{\label{fig:mpw_tow74}
The ADC count distribution of the data can be fitted well with
a falling exponential curve (top).
The absolute slope values decrease with increasing luminosity (bottom).
Data are shown for Tower\#74 of the West MP.}
\end{figure}

The first data from Run~II have been used to calibrate and commission the MPs.
Although a precise energy calibration of the MP is not crucial to the
understanding of diffractive processes, an attempt was made 
to estimate the energy scale of jets and particles.
To this end, a Monte Carlo simulation
was used to calibrate the pseudorapidity dependence of the particles' energies and thereby 
the tower-by-tower relative response.
For each tower, the ADC count distribution of the data can be fitted 
well with a falling exponential curve, as shown in Figure~\ref{fig:mpw_tow74}~(top).

Due to pile-up effects at larger rapidity regions, 
a luminosity dependence of the
ADC count distribution slope can be observed (Fig.~\ref{fig:mpw_tow74}, bottom). 
A linear fit seems to describe well the data with the slope decreasing with 
increasing values of the instantaneuos luminosity. As expected, lower slope 
values are measured for the inner rings, where the particles' energies are larger.
The distribution of the residuals 
is well represented by a Gaussian with a r.m.s. of
approximately 7\% of the distribution mean for all MP rings.
In order to estimate the energy calibration, the ADC distributions are compared
with a Monte Carlo simulation for a sample of minimum bias events.
The slopes are first equalized separately in each ring and are then adjusted to the 
slopes predicted from Monte Carlo for different pseudorapidity regions.

\begin{figure}[htb]
\includegraphics*[width=\hsize]{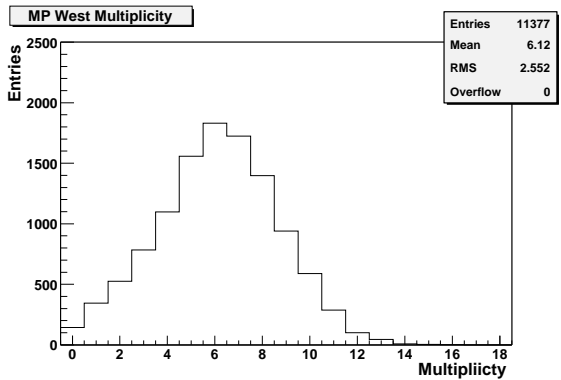}
\vspace*{-1.5cm}
\caption{\label{fig:mpw_mult_and_et}
Particle multiplicity in West MP.}
\end{figure}

\begin{figure}[htb]
\includegraphics*[width=\hsize]{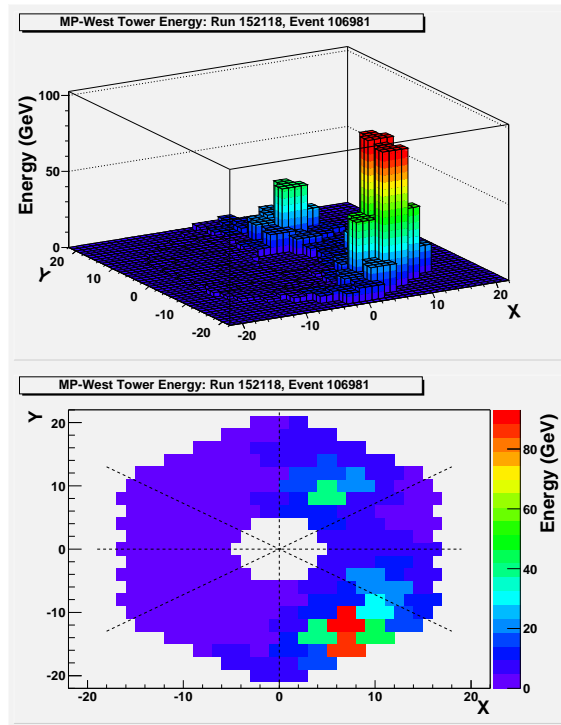}
\vspace*{-1.5cm}
\caption{\label{fig:mp_2jet}
Event display of a 2-``jet'' event in the East MP.
The vertical axis shows the signal pulse height
measured in units of GeV.
The term ``jet'' is used to indicate a hadron or
electromagnetic shower and not an actual jet of particles.
The 2-dimensional plot (bottom) shows the {\it x-y} 
coordinate of the particles hitting the MP.}
\end{figure}

The particle multiplicity is measured by counting the number of ``peaks'' 
(signal above detector noise) or {\em seed} towers.
Figure~\ref{fig:mpw_mult_and_et} 
shows the particle multiplicity in the West MP in a minimum bias event sample, 
which is in agreement with Monte Carlo expectations.
Typically, one peak corresponds to one particle. 
When more particles are confined to the same $\eta - \phi$ region,
the energy measured is proportionally larger.
In order to correctly calibrate the tower energy scale, the lateral spread of the 
shower was measured.
About 25\% of the energy of the shower is deposited in the seed tower. 
The remainder is deposited in the surrounding towers.
Particles with large energies in the MP only correspond to small
transverse energies ($E\approx E_T \cdot \theta_{MP}$, where $\theta_{MP}\approx 30\div 50$~mrad).
Figure~\ref{fig:mp_2jet} shows an event display with two jets in the MP
(the term ``jet'' is used here to indicate a cluster of towers, which
is most likely due to one particle interacting in the MP).
This event was selected from a minimum bias data sample.

\section{PROSPECTS}
The program for diffractive physics during Run~II at the Tevatron includes
studies of soft and hard diffraction and of double pomeron exchange.
The Forward Detectors are an essential component of this program 
at the CDF experiment.
The MP calorimeters are needed to measure 
the flow of the event energy in the very forward rapidity region.

The detectors have been installed, the study of their performance and some 
of the results obtained have been discussed here.
The MPs are presently collecting good quality data 
for further exploring difractive physics.

In conclusion, after many years of preparation, Run~II is finally becoming a reality.

\section*{ACKNOWLEDGMENTS}
My warm thanks to the organizers for a wonderful workshop with a special taste.
Also to the {\it contradaioli} of {\it Civetta} and {\it Leocorno} 
for great hospitality.

\end{document}